\documentstyle[aps,epsfig,floats]{revtex}
\newcommand{\bm}[1]{\mbox{\boldmath $#1$}}
\newcommand{\be}{\begin{equation}}
\newcommand{\ee}{\end{equation}}
\newcommand{\bea}{\begin{eqnarray}}
\newcommand{\eea}{\end{eqnarray}}

\newcommand{\bfp}{\mbox{\boldmath $p$}}
\newcommand{\NP}[1]{{\it Nucl.\ Phys.}\ {\bf #1}}
\newcommand{\ZP}[1]{{\it Z.\ Phys.}\ {\bf #1}}
\newcommand{\PL}[1]{{\it Phys.\ Lett.}\ {\bf #1}}
\newcommand{\PR}[1]{{\it Phys.\ Rev.}\ {\bf #1}}
\newcommand{\PRL}[1]{{\it Phys.\ Rev.\ Lett.}\ {\bf #1}}

\newcommand{\SNP}[1]{{\it Sov.\ J.\ Nucl.\ Phys.}\ {\bf #1}}
\newcommand{\EPJ}[1]{{\it Eur.\ Phys.\ J.}\ {\bf #1}}
\newcommand{\IJMP}[1]{{\it Int.\ J.\ Mod.\ Phys.}\ {\bf #1}}
\begin{document}
\tighten
\thispagestyle{empty}
\title{
\begin{flushright}
\begin{minipage}{4 cm}
\small
hep-ph/9911207\\ 
VUTH 99-24
\end{minipage}
\end{flushright}
\vspace{5mm}
Reassessment of the Collins Mechanism for Single-spin Asymmetries and 
the behavior of \bm{\Delta d(x)} at large \bm{x}.
\protect} 
\vspace{5mm}
\author{M. Boglione and E. Leader
\footnote{Permanent address: Theoretical Physics Research Unit, 
Birkbeck College, University of London, Malet Street, 
London WC1E 7HX, U.K.}
\vspace{5mm}
\mbox{}\\
{\it Division of Physics and Astronomy, Faculty of Science, Vrije 
Universiteit Amsterdam}\\
{\it De Boelelaan 1081, NL-1081 HV Amsterdam, The Netherlands}}

\maketitle

\vspace{1cm}

\begin{abstract}
It is shown that the Collins mechanism explanation of the transverse 
single-spin asymmetries in $p^{\uparrow}p \to \pi X$ leads to a transversely 
polarized $d$ quark density $\Delta _T d(x)$ which violates the Soffer bound 
when one uses several standard forms for the longitudinally polarized $d$ 
quark density $\Delta d(x)$ obtained from polarized deep inelastic scattering. 
Imposition of the Soffer bound with these $\Delta d(x)$ yields results in 
hopeless disagreement with the data. Remarkably, imposition of the Soffer 
bound, but using parametrizations of $\Delta d(x)$ that respect the PQCD 
condition $\Delta q(x)/q(x) \to 1 $ as $x \to 1$, leads to an excellent fit 
to most of the data. The implications for the polarized DIS neutron 
longitudinal asymmetry $A_1^n$ at large $x$ are dramatic.
\end{abstract}

\vspace{0.6cm}

~~~~~~~~~~PACS numbers: 13.60.Hb, 13.85.Ni, 13.87.Fh, 13.88.+e
  
\section{Introduction}

One of the major challenges to the QCD-parton model is the explanation
of the large (20-40\%) single-spin  asymmetries found in many
semi-inclusive hadron-hadron reactions,  of which the most dramatic are
the polarization of the lambda in $pp \to \Lambda X$ and the asymmetry under
reversal of the transverse spin of the proton in $p^{\uparrow} p \to \pi X$. 
The challenge arises from the fact that in the standard approach the basic
``two parton $\to$ two parton'' reactions involved in the perturbatively
treated  hard part of the scattering do not possess this kind of
asymmetry.

Already some time ago, Efremov and Teryaev \cite{et} suggested  a mechanism
for these asymmetries utilizing ``three parton $\to$ two parton'' amplitudes
for the hard scattering. This, however, necessitates the introduction of
a new unknown soft two-parton density, namely the correlated probability
of finding in the polarized proton a quark with momentum fraction $x_1$
and a gluon with fraction $x_2$. This quark-gluon correlator contains the
dependence on the transverse spin of the proton. A fully consistent 
application of the approach has not yet been carried out, though a 
significant step in the direction has been taken recently by Sterman 
and Qiu \cite{sq}.

Some time ago Sivers \cite{siv} and Collins \cite{col} suggested 
mechanisms for the asymmetries, which are within the framework of the 
standard ``two parton $\to$ two parton'' picture, but in which the 
transverse momentum ${\bf k_T}$ of the quark in the hadron plays an 
essential role. 
Sivers introduces a parton density which depends on the transverse spin 
${\bf S_T}$ of the proton
in the form  ${\bf S_T \cdot (k_T \times p)}$ where ${\bf p}$ is the 
momentum of the
polarized proton. However such a structure is forbidden by time reversal
invariance \cite{col}. Nonetheless, on the grounds that the time-reversal
argument could be invalidated by initial or final state interactions,
Anselmino, Boglione and Murgia \cite{abm95,abm99} have applied the Sivers 
mechanism to 
$p^{\uparrow}p \to \pi X$ and shown that a very good fit to the data at 
large $x_F$ can
be achieved. The problem is that this approach raises a fundamental
question: if the initial and final state interactions are important then
it is hard to see why the underlying factorization into hard and soft
parts is valid. (The advantage of the quark-gluon correlator approach is
that it effectively provides initial and final state interactions calculable 
at the parton level). 
In the Collins mechanism  the asymmetry arises in the
fragmentation of a polarized quark into a pion with transverse momentum
$p_T$, described by a fragmentation function of the form  
$\Delta ^N D(z,p_T)$  (see Section II) which is
convoluted with the transverse spin dependent parton density $\Delta _T q$,
about which nothing is known experimentally. This structure is not in
conflict with time reversal invariance.
The definition of transverse spin dependent parton densities and possible
methods of determining them have been studied by Artru and Mekhfi~\cite{am90},
Cortes, Pire and Ralstone~\cite{cpr} and Jaffe and Ji~\cite{jj}. 
Note that in the latter they are referred to as "transversity distributions" 
and that notation differs amongst all these papers.

An estimate of the size of the Collins effect was first made by Artru, 
Czyzewski and Yabuki \cite{acy97}, but more recently 
Anselmino, Boglione and Murgia \cite{abm99} have demonstrated that an 
excellent fit to the data on $p^{\uparrow} p \to \pi X$ can be obtained 
with the Collins mechanism.
However their fit  is problematic  since the $\Delta _T q$'s 
 used in the fit violate the Soffer bound \cite{sof}
\be
|\Delta _T q(x)| \le \frac{1}{2} [ q(x) + \Delta q(x)] 
\label{Soff}
\ee
at large $x$ when $\Delta q(x)$, the usual longitudinal polarized parton 
density, is taken from any of the standard parametrizations of the 
longitudinal and unpolarized densities: 
\begin{itemize}
\item{GRSV-GRV = Gl\"uck, Reya, Stratmann and Vogelsang \cite{grsv} + 
Gl\"uck, Reya, Vogt \cite{grv}},
\item{GS-GRV = Gehrmann and Stirling \cite{gs} + 
Gl\"uck, Reya, Vogt \cite{grv}},
\item{LSS-MRST = Leader, Sidorov and Stamenov \cite{lss} + 
Martin, Roberts, Stirling and Thorne \cite{mrst}}.
\end{itemize}
Note that in the above parametrizations the polarized densities are linked 
to particular parametrizations of the unpolarized densities, as indicated.

The key point is that the $\pi^-$ data demand a
large magnitude for $\Delta _T d$ at large $x$, whereas $\Delta d(x)$ is almost
universally taken negative for all $x$, thereby making the Soffer bound
much more restrictive for the $d$ than  for the $u$ quark.

This raises an intriguing question. There is an  old perturbative QCD
argument \cite{fj} that strictly in the limit $x \to 1$
\be
\frac{\Delta q(x)}{q(x)} \to 1\;,
\ee
which would imply that $\Delta d(x)$ has to change sign and become positive 
at large $x$. 
The polarized DIS data certainly require a negative $\Delta d(x)$ in the range
$0.004 < x \leq 0.75$ where data exist, but there is no reason why 
$\Delta d(x)$ should not change sign near or beyond $0.75$. Indeed there is a 
parameterization of the parton densities by Brodsky, Burkhardt and
Schmidt [BBS] \cite{bbs} which has this feature built into it.
The original BBS fit is not really competitive  since evolution in $Q^2$
was not taken into account, but a proper QCD fit based on the BBS
parameterization was shown by Leader, Sidorov and Stamenov \cite{lss-bbs} 
to give an adequate fit to the polarized DIS data.

In this paper we address the question of the correct use of the Collins 
mechanism in which the Soffer bound is respected. We find that it is
impossible to get a good fit to the $\pi^{\pm}$ data when the
magnitude of $\Delta _T d$ is controlled by (1) in which $\Delta d(x)$ from 
any of the standard forms given above is used. 
On the contrary, and most surprisingly, we find that
 parametrizations in which $\Delta d(x)/d(x) \to 1$ as $x \to 1$ allow a 
$\Delta _T d(x)$ that leads to an excellent fit to most of the pion data.

In Section II we briefly describe the Collins mechanism, and present our 
results in Section III. Conclusions follow in Section IV.

\section{The Model}

As mentioned in the Introduction, we require the Soffer inequality, 
Eq.(\ref{Soff}), to 
be respected by the distribution functions $\Delta _T u$  and $\Delta _T d$  
determined by our fit.
Besides, the positivity constraint $\Delta ^N D (z) \le 2\,D(z)$ must hold, 
since $D(z)=\frac{1}{2} \, [D^{\uparrow}(z) + D^{\downarrow}(z)]$ and 
$\Delta ^N D (z)=[D^{\uparrow}(z) - D^{\downarrow}(z)]$.
Therefore, the parametrizations are set so that these conditions are 
automatically fulfilled, in the following way. 
First we build a simple function of the form $x^a (1-x)^b$ (or
$z^\alpha (1-z)^\beta$, as appropriate), where the powers $a,\,b$ or 
$\alpha,\,\beta$ are $\geq 0$, and we divide by their maximum value. 
By allowing an extra multiplicative 
constant factor to vary from $-1$ to $1$, we obtain a function which, 
in modulus, will never be larger than 1. Then we parameterize $\Delta _T q(x)$ 
and $\Delta ^N D(z)$ by multiplying the functions  we built by the constraint 
given by the Soffer inequality or the positivity limit. In this way we make 
sure that the bounds are never broken.

For the transversity distribution functions we set
\be
\Delta _T u(x) =  N_u \; \frac{x^a (1-x)^b}{\frac{a^a b^b}{(a+b)^{a+b}}} \;
            \Bigl\{\frac{1}{2} \, [ u(x) + \Delta u(x)] \Bigr\},
\ee
\be
\Delta _T d(x)  =  N_d \; \frac{x^{c} (1-x)^{d}}
            {\frac{{c}^{c} d^{d}}{(c+d)^{c+d}}} \;
            \Bigl\{\frac{1}{2} \, [ d(x) + \Delta d(x)] \Bigr\},
\ee
with $|N_{u,d}| \le 1$. Here $q(x)$ and $\Delta q(x)$ are the whole 
distribution functions, i.e. they contain valence and sea contributions 
(but this is irrelevant at large $x$ since there the contribution of the 
sea is negligible).
As in the previous calculation, only $u$ and $d$ contributions are taken into 
account in the polarized proton, so that 
\be
\Delta _T \overline u(x) = 
\Delta _T \overline d(x) =  \Delta _T s (x) = \Delta _T \overline s (x) = 0.
\ee

For the functions $q(x)$ and  $\Delta q(x)$ we use, for comparison, the 
``standard'' parton parametrizations mentioned in Section I and two further 
parametrizations, one due to Brodsky, Burkhardt and Schmidt (BBS) \cite{bbs} 
which ignores $Q^2$-evolution, and a more consistent version of this, due 
to Leader, Sidorov and Stamenov (LSS)$_{BBS}$ \cite{lss-bbs} which includes 
the $Q^2$-evolution. 
These will be explained in more detail in Section III. 

For the fragmentation function we have
\be
\Delta ^N D (z) = N_F \; \frac{z^\alpha (1-z)^\beta}
                 {\frac{\alpha^\alpha \beta^\beta}{(\alpha+\beta)^
                 {\alpha+\beta}}} \;
                 \Bigl[ \,2\,D(z)\, \Bigr],
\ee
with $|N_F| \le 1$. Here we take into account only valence contributions, 
so that isospin symmetry and charge conjugation give
\be
\Delta ^N D^u_{\pi^+} = \Delta ^N D^d_{\pi^-} = \Delta ^N D (z) \, ,
\ee
\be 
\Delta ^N D^{\overline u}_{\pi^+} = \Delta ^N D^{\overline d}_{\pi^-} = 0
\ee
and 
\be
\Delta ^N D^u_{\pi^0} = \Delta ^N D^d_{\pi^0} = 
\Delta ^N D^{\overline u}_{\pi^0} = \Delta ^N D^{\overline d}_{\pi^0} =
\frac{1}{2} \,\Delta ^N D (z). 
\ee
Notice that $\Delta ^N D$ is, in fact, a function of the intrinsic 
transverse momentum $\bfp _T$. $\Delta ^N D_{h/a^{\uparrow}}(z,\bfp_T)$ is 
defined as the difference between the number density of hadrons $h$, 
a pion in our case, with longitudinal momentum fraction $z$ and transverse 
momentum $\bfp _T$, originating from a transversely polarized parton $a$ with 
spin either $\uparrow$ or $\downarrow$, respectively
\bea
\Delta^N D_{h/a^{\uparrow}}(z, \bfp_T) &\equiv&
\hat D_{h/a^{\uparrow}}(z, \bfp_T) - \hat D_{h/a^{\downarrow}}(z, \bfp_T) 
\label{delf1} \nonumber \\
&=& \hat D_{h/a^{\uparrow}}(z, \bfp_T)-\hat D_{h/a^{\uparrow}}(z, - \bfp_T) \>,
\eea
where the second line follows from the first one by rotational invariance.
Details on the integration over the transverse momentum ${\bfp_T}$ and its  
dependence on $z$ are given in Ref.~\cite{abm99} (see Eqs. (17) and (19)). 
The unpolarized fragmentation function for pions, $D(z)$, is taken from 
Binnewies et al. \cite{bkk1}.

With these ingredients we are now ready to calculate, in complete analogy 
with Ref. \cite{abm99}, the $p^{\uparrow}p \to \pi X$ single spin asymmetry 
\be
A_N = \frac{d\sigma ^{\uparrow} - d\sigma ^{\downarrow}}
           {d\sigma ^{\uparrow} + d\sigma ^{\downarrow}} \, .
\ee 
Here
\bea
d\sigma^{\uparrow} - d\sigma^{\downarrow} &=& 
\sum_{a,b,c,d} \int \frac{dx_a \, dx_b}{\pi z} \, d^2 {\bfp_T}
\label{dscol} \nonumber \\ 
&\times& \Delta _T q^{a}(x_a) \> q^{b}(x_b) \>
\Delta_{NN} \hat\sigma^{ab \to cd}(x_a, x_b, {\bfp_T}) \>
\Delta^N D_{\pi/c}(z, {\bfp_T}) \, ,
\eea
where
\be
\Delta_{NN} \hat\sigma^{ab \to cd} = 
\frac{d \hat \sigma ^{a^{\uparrow}b \to c^{\uparrow}d}}{d \hat t} -
\frac{d \hat \sigma ^{a^{\uparrow}b \to c^{\downarrow}d}}{d \hat t} \, ,
\ee
and
\bea
d\sigma^{\uparrow} + d\sigma^{\downarrow} &=& 2\, d\sigma^{unp} = 
2 \sum_{a,b,c,d} \int \frac{dx_a \, dx_b}{\pi z} 
\label{dsunp} \\ 
&\times& q^a(x_a) \> q^b(x_b) \>
\frac{d\hat\sigma^{ab \to cd}}{d\hat t}(x_a, x_b) \> D_{\pi/c}(z) \>.\nonumber
\eea
All details about the calculation can be found in Ref. \cite{abm99}.
The relation between the above  notation and that of \cite{abm99} is : 
$q^a(x) = f_{a/p}(x)$ and 
$\Delta _T q^a(x) = P^{a/p^{\uparrow}}\,f_{a/p}(x)$.

\section{Results}

We start by running two fits to the $E704$ experimental data \cite{e704} 
using the popular GS \cite{gs} polarized densities in conjunction with the GRV 
\cite{grv} unpolarized densities, and the latest LSS \cite{lss} polarized densities in conjunction 
with the MRST \cite{mrst} unpolarized densities. 
It should be noted that the 1996 analysis of Gehrmann and Stirling was done 
prior to the publication of a great deal of new, high precision data on 
polarized DIS, whereas the Leader, Sidorov and Stamenov analysis includes all 
of the present world data.
Fig.~1 shows the complete failure to fit the data when the Soffer bound is 
implemented using the GS-GRV parametrizations ($\chi ^2/DOF = 25$ !).
The corresponding transverse densities $\Delta _T u (x)$ and $\Delta _T d (x)$ 
are shown in Fig.~2. In this fit only $|N_F|=1$ is possible and Fig.~2 
corresponds to $N_F=-1$. The sign is discussed later. 

A somewhat better picture emerges when using the LSS-MRST results to implement 
the Soffer bound: Fig.~3. The fit looks reasonable out to $x_F=0.4$ but fails 
beyond that. One finds $\chi ^2/DOF = 6.12$.  
The transverse densities are shown in Fig.~4, where the curves correspond to 
negative $N_F$.

Note that in both these fits one finds $\alpha=\beta=0$ in Eqn.~(6), showing 
that the magnitude of $\Delta ^N D (z)$ is maximized at each $z$-value.

The reason for the failure of GS-GRV case and for the relative success of 
LSS-MRST can be understood by observing in Fig.~5 that the Soffer bound 
on $\Delta _T d (x)$ is {\em much} more restrictive at large $x$ in the GS-GRV 
case. Comparison of Fig.~5 with Fig.~6 also indicates the source of the 
problem. The asymmetries for $\pi^{\pm}$ are of roughly equal magnitude 
whereas the Soffer bound restrictions are much more severe for the $d$ quark 
as a consequence of $\Delta d (x)$ being negative for all $x$.

This suggests an intriguing possibility. The polarized DIS data only exist for 
$x \le 0.75$ and there is really very little constraint from DIS on the 
$\Delta q (x)$ for $x$ near to and beyond this value. At the same time there 
are perturbative QCD arguments  \cite{fj} which suggest that 
\be
\frac{\Delta q (x)}{q(x)} \to 1 \;\;\; {\rm as} \;\;\; x \to 1
\label{deltaq/q}
\ee
and, indeed, even more precisely, that
\be
q(x) - \Delta q (x) \; \propto \; (1-x)^2\,q(x) \;\;\; {\rm as} \;\;\; 
x \to 1 \;.
\label{deltaq-q}
\ee
This constraint is almost universally ignored in parameterizing the 
$\Delta q (x)$, on the grounds that (\ref{deltaq-q}) is incompatible with 
the evolution equations. But this is a ``red herring'' since the  
evolution equations do not hold in the region where (\ref{deltaq-q}) is 
valid, approaching the border of the exclusive region.

The imposition  of (\ref{deltaq/q}) is exactly what we need for $\Delta q(x)$ 
to change sign and become positive at large $x$, thereby diminishing the 
restrictive power of the Soffer bound on $\Delta _T d(x)$.

In fact there does exist a parametrizations of the $\Delta q(x)$ which respects 
(\ref{deltaq/q}) and (\ref{deltaq-q}), namely that of Brodsky, Burkhardt 
and Schmidt (BBS) \cite{bbs}. Unfortunately BBS did not include any 
$Q^2$-evolution where determining the numerical values of their parameters 
from the DIS data, so their fit is not really adequate.

However, Leader, Sidorov and Stamenov \cite{lss-bbs} made an extensive study, 
using the BBS 
functional forms, but including $Q^2$-evolution, and found a very good fit 
(LSS)$_{BBS}$ to the polarized DIS data available in 1997.

It can be seen in Fig.~5 that the Soffer bound on $\Delta _T d(x)$ is much 
less restrictive for the BBS case and that the (LSS)$_{BBS}$ bound is rather 
similar to that of the LSS-MRST case, but is less restrictive for $x \ge 0.7$.
It is important to realize that although the $\Delta _T d(x_a)$ needed in 
(\ref{dscol}) are tiny for such large values of $x_a$, this is compensated 
for by the 
fact that large $x_F$ then demands very small $x_b$, where the unpolarized 
densities grow very large.

%%%%%%%%%%%%%%%%%%%%%%%%%%%%%%%%%%%%%%%%%%%%%%%%%%%%%%%%%%%%%%%%%%%%%%%
\begin{table}[ht]
\begin{tabular}{c c c c c } 
\rule[-0.2cm]{0cm}{2mm} \rule[0.2cm]{0cm}{2mm}
~&GS-GRV       & LSS-MRST&  BBS    & (LSS)$_{BBS}$ \\ 
\hline \rule[0.2cm]{0cm}{2mm}
~~ $N_F\,N_u$ &$-0.43~~$& $-0.73~~$& $-0.54~~$& $-0.49~~$ \\ 
~~ $N_F\,N_d$ &$1.00$   & $1.00$   & $0.88$   & $0.91$    \\ 
~~  $a$       &$4.33$   & $3.03$   & $3.17$   & $3.46$    \\ 
~~  $b$       &$0.00$   & $0.00$   & $0.00$   & $0.00$    \\ 
~~  $c$       &$0.00$   & $3.48$   & $3.57$   & $3.32$    \\ 
~~  $d$       &$0.00$   & $0.76$   & $0.00$   & $0.00$    \\ 
$\chi^2/DOF$  &$ 25 $   & $6.12$   & $1.45$   & $2.41$ 
\rule[-0.2cm]{0cm}{2mm}
\end{tabular}
\vspace{0.5cm}
\caption{Parameters determined by the fit in the four different 
parameterization schemes and the corresponding $\chi^2/DOF$.}
\end{table}
%%%%%%%%%%%%%%%%%%%%%%%%%%%%%%%%%%%%%%%%%%%%%%%%%%%%%%%%%%%%%%%%%%%%%%%
%

Note that the relative signs of $N_u$ and $N_d$ are opposite, but their 
absolute  signs are not determined since, in principle, $N_F$ can be positive 
or negative. However, if one uses an $SU(6)_F$ wave function for the proton, 
one finds $\Delta _T u$ positive and $\Delta _T d$ negative, so it seems 
reasonable to hypothesize that $N_u > 0$ and $N_d < 0$.
For this reason we have chosen $N_F$ to be negative in the above.
Note that $N_u$ and $N_d$ are not a direct measure of the magnitudes of 
$\Delta _T u$ and $\Delta _T d$. Their role is linked specifically to the 
Soffer bound. The relative behavior of $\Delta _T u$ and $\Delta _T d$ can 
be seen in Figs.~2,~4,~9,~10.

Indeed, as expected, we find that a significantly better fit to the asymmetry 
data is achieved using the BBS and the (LSS)$_{BBS}$ parametrizations, with 
$\chi ^2/DOF = 1.45$ and $\chi ^2/DOF = 2.41$ respectively. As can be seen in  
Fig.~7~and~8. The curves reproduce the trends in the data right out to 
$x_F \sim 0.7$. Figs.~9~and~10 show how similar the allowed ranges of 
transverse polarized densities are in the two cases. 
In Fig.~9, $~0.88\le |N_F| \le 1~$, whereas in Fig.~10 
$~0.91 \le |N_F| \le 1~$. 
As before the curves correspond to negative $N_F$.

The parameter values for all the parametrizations are shown in Table 1, 
where it should be recalled $|N_F|$, $|N_u|$ and $|N_d|\;\le\; 1$.

\section{Conclusions}

We have demonstrated that the Collins mechanism is able to explain much of 
the data on the transverse single spin asymmetries in 
$p^{\uparrow} p \to \pi X$, namely the data in the region $x_F \leq 0.7$, if, 
and only if, the longitudinal polarized $d$-quark density, which is negative 
for small and moderate $x$, changes sign and becomes positive at large $x$. 
There is hopeless disagreement when using the longitudinal polarized densities 
due to Gehrmann and Stirling \cite{gs}, and matters are significantly better 
when using the most up to date parameterization of Leader, Sidorov and 
Stamenov \cite{lss}. But the most successful fits arise from parametrizations 
\cite{bbs,lss-bbs} which respect the PQCD condition $\Delta d(x)/d(x) \to 1$ 
as $x \to 1$.

For parametrizations of $\Delta d(x)$ with this property there are 
interesting consequences in polarized DIS, namely, the neutron longitudinal 
asymmetry  $A_1^n(x)$ should change sign and tend to $1$ as $x \to 1$  
(see Fig.~11). The region of large $x$ has hardly been explored in polarized 
DIS up to the present. Clearly a study of this region might turn out to be 
very interesting.

There remains, however, the problem of the $p^{\uparrow} p \to \pi X$ data at 
the largest values of $x_F$ so far measured, i.e. $0.7\leq x_F \leq 0.82$.
It does not seem possible to account for these asymmetries within the 
framework of the Collins mechanism.
On the other hand Qiu and Sterman \cite{sq}, using a ``three parton $\to$ two 
parton'' amplitude for the hard partonic scattering and a ``gluonic pole'' 
mechanism, claim to be able to reproduce the very large asymmetries at 
$x_F\sim 0.8$.  However their study must be considered as preliminary, since 
it relies on a completely {\it ad hoc} assumption that the essential new 
twist-three quark-gluon-quark correlator function $T_F^{(v)}(x,x)$, for 
given flavour $f$, is proportional to $q_f(x)$, and no attempt is made to 
fit the detailed $x_F$-dependence of the data.

\vspace{0.7cm}

\begin{center}{\small \bf ACKNOWLEDGMENTS}\end{center}

The authors are grateful to M. Anselmino and F. Murgia for general comments, 
and to S. Brodsky, R. Jakob, P. Kroll and I. Schmidt for discussions 
concerning the condition $\Delta q(x)/q(x) \to 1$ as $x \to 1$. 
E. L. is grateful for the hospitality of the Division of Physics and 
Astronomy of the Vrije Universiteit, Amsterdam.
This research project was 
supported by the Foundation for Fundamental Research on Matter (FOM) and the 
Dutch Organization for Scientific Research (NWO).

\newpage

%%%%%%%%%%%%%%%%%%%%%%%%%%%%%%%%%%%%%%%%%%%%%%%%%%%%%%%%%%%%%%%%%%%%%%%%%%
\begin{figure}[t]
\begin{center}
\mbox{~\epsfig{file=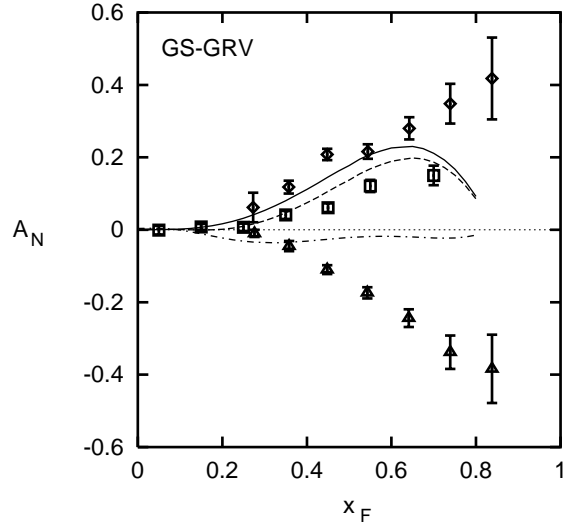,angle=-90,width=8cm}} 
\vspace{0.4cm} 
\caption{
Single spin asymmetry for pion production in the process 
$p^{\uparrow} p \to \pi X$ as a function of $x_F$, 
obtained by using the GS-GRV \protect \cite{grv,gs} sets of distribution 
functions.
The solid line refers to $\pi^+$, the dashed line to 
$\pi^0$ and the dash-dotted line to $\pi^-$. }
\end{center}
\end{figure}
%%%%%%%%%%%%%%%%%%%%%%%%%%%%%%%%%%%%%%%%%%%%%%%%%%%%%%%%%%%%%%%%%%%%%%%%%%%
%
%%%%%%%%%%%%%%%%%%%%%%%%%%%%%%%%%%%%%%%%%%%%%%%%%%%%%%%%%%%%%%%%%%%%%%%%%%
\begin{figure}[b]
\begin{center}
\mbox{~\epsfig{file=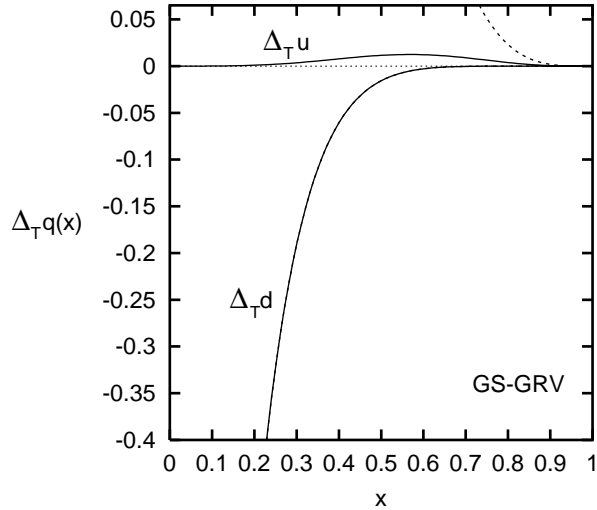,angle=-90,width=8cm}}
\vspace{0.4cm} 
\caption{
The distribution functions $\Delta _T u(x)$ and 
$\Delta _T d(x)$, as obtained by using the GS-GRV \protect \cite{grv,gs} 
distribution 
functions. The dotted lines are the 
boundaries imposed by the Soffer inequality.  
For $\Delta _T d(x)$ the dotted line is invisible since $\Delta _T d(x)$ 
completely saturates the Soffer bound in the whole $x$ region. For discussion 
of signs, see text.}
\end{center}
\end{figure}
%%%%%%%%%%%%%%%%%%%%%%%%%%%%%%%%%%%%%%%%%%%%%%%%%%%%%%%%%%%%%%%%%%%%%%%%%%
%
%%%%%%%%%%%%%%%%%%%%%%%%%%%%%%%%%%%%%%%%%%%%%%%%%%%%%%%%%%%%%%%%%%%%%%%%%%
\begin{figure}[t]
\begin{center}
\mbox{~\epsfig{file=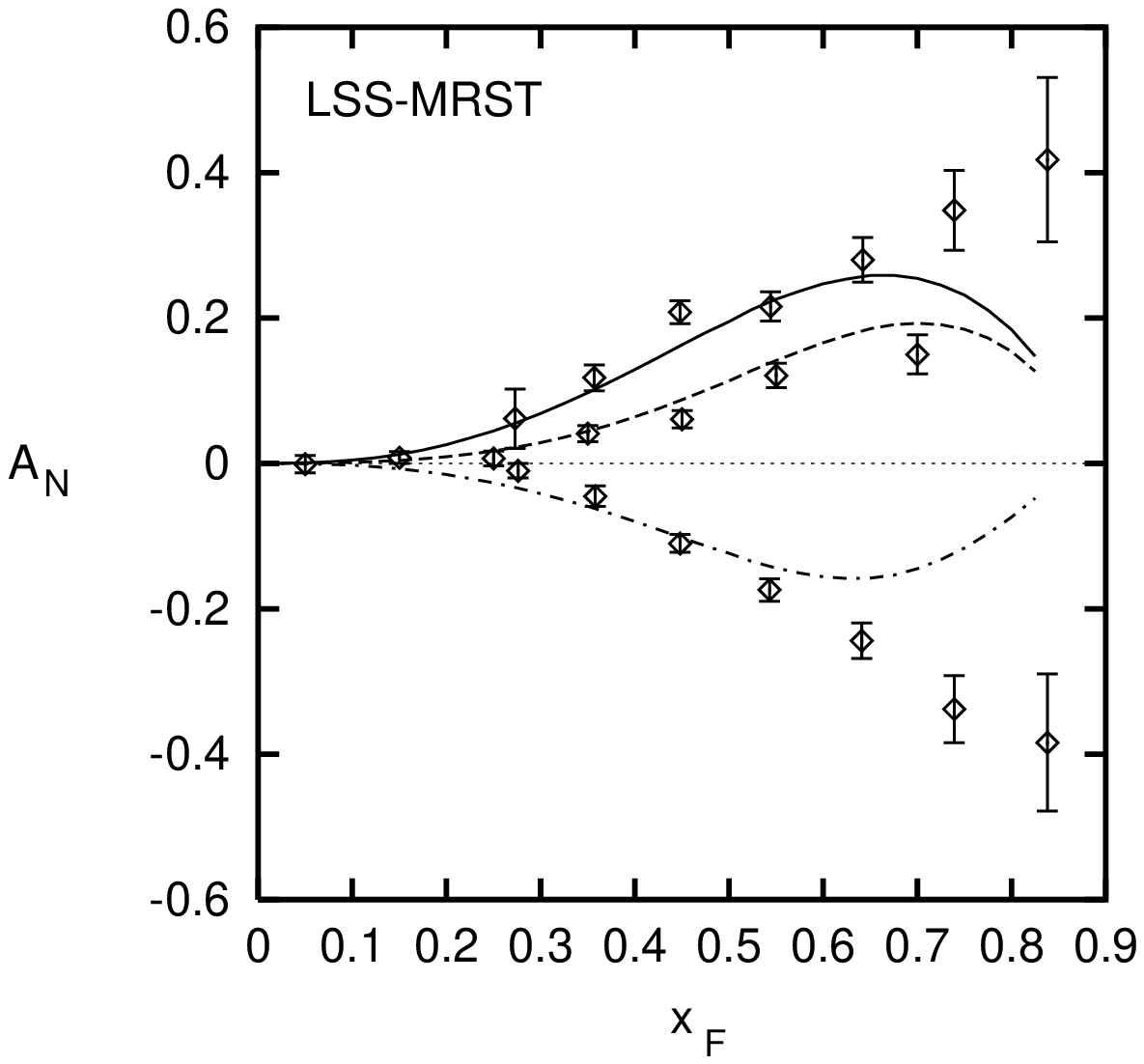,angle=-90,width=8.0cm}}
\vspace{0.4cm}
\caption{
The single spin asymmetry for pion production in the process 
$p^{\uparrow} p \to \pi X$ as a function of $x_F$, obtained  using the 
LSS-MRST \protect \cite{lss,mrst} distribution functions. 
The solid line refers to $\pi^+$, the dashed line to 
$\pi^0$ and the dash-dotted line to $\pi^-$.}
\end{center}
\end{figure}
%%%%%%%%%%%%%%%%%%%%%%%%%%%%%%%%%%%%%%%%%%%%%%%%%%%%%%%%%%%%%%%%%%%%%%%%%%
%
%%%%%%%%%%%%%%%%%%%%%%%%%%%%%%%%%%%%%%%%%%%%%%%%%%%%%%%%%%%%%%%%%%%%%%%%%%
\begin{figure}[b]
\begin{center}
\mbox{~\epsfig{file=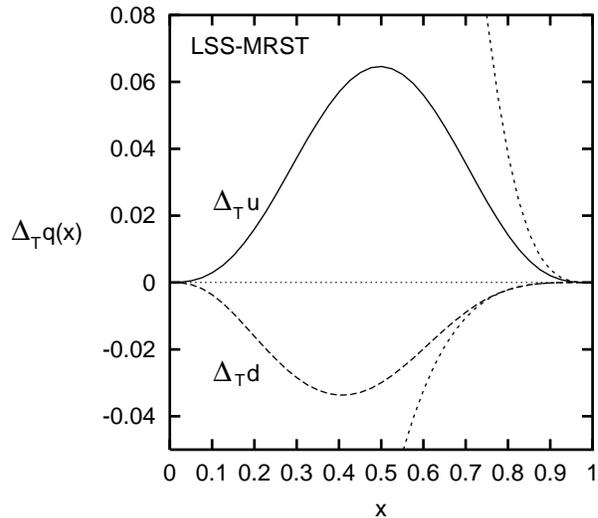,angle=-90,width=8.0cm}} 
\vspace{0.4cm}
\caption{
The distribution functions $\Delta _T u(x)$ and 
$\Delta _T d(x)$ versus $x$, 
as determined by the fit using the LSS-MRST \protect \cite{lss,mrst} 
distribution functions.  The dotted lines are the 
boundaries imposed by the Soffer inequality. For discussion of signs 
see text.}
\end{center}
\end{figure}
%%%%%%%%%%%%%%%%%%%%%%%%%%%%%%%%%%%%%%%%%%%%%%%%%%%%%%%%%%%%%%%%%%%%%%%%%%
%
%%%%%%%%%%%%%%%%%%%%%%%%%%%%%%%%%%%%%%%%%%%%%%%%%%%%%%%%%%%%%%%%%%%%%%%%%%
\begin{figure}
\begin{center}
\mbox{~\epsfig{file=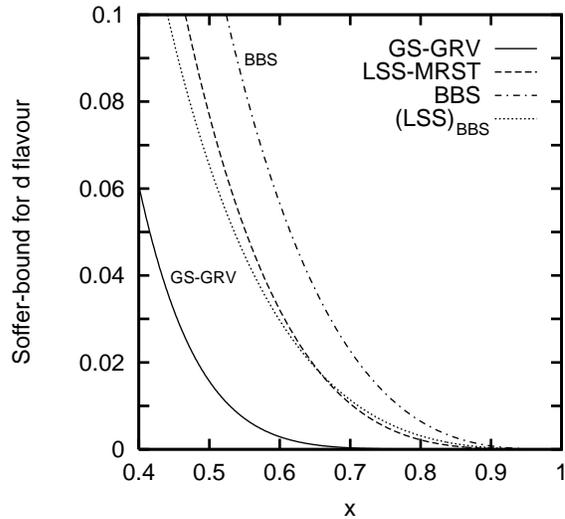,angle=-90,width=8cm}}
\vspace{0.4cm}
\caption{
The boundaries imposed on $\Delta _T d$ by the Soffer inequality. 
Note that the GS-GRV distribution functions give a 
much tighter bound than any of the other parameterization sets.}
\end{center}
\end{figure}
%%%%%%%%%%%%%%%%%%%%%%%%%%%%%%%%%%%%%%%%%%%%%%%%%%%%%%%%%%%%%%%%%%%%%%%%%%%
%
%%%%%%%%%%%%%%%%%%%%%%%%%%%%%%%%%%%%%%%%%%%%%%%%%%%%%%%%%%%%%%%%%%%%%%%%%%
\begin{figure}
\begin{center}
\mbox{~\epsfig{file=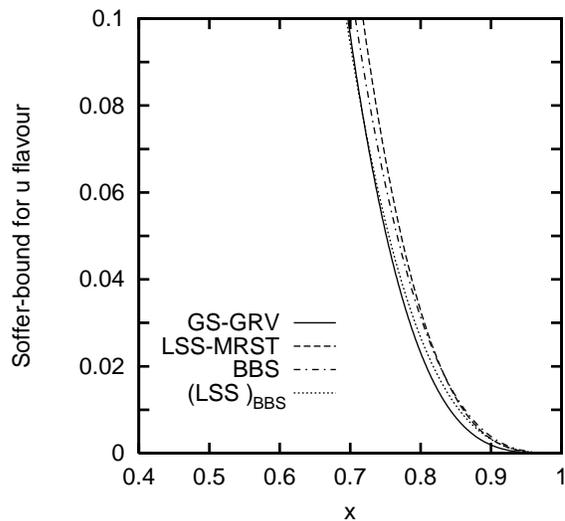,angle=-90,width=8cm}}
\vspace{0.4cm}
\caption{
The boundaries imposed on $\Delta _T u$ by the Soffer inequality.
For the $u$ flavour the Soffer bound is very similar in each 
set of parametrizations.}
\end{center}
\end{figure}
%%%%%%%%%%%%%%%%%%%%%%%%%%%%%%%%%%%%%%%%%%%%%%%%%%%%%%%%%%%%%%%%%%%%%%%%%%%
%
%%%%%%%%%%%%%%%%%%%%%%%%%%%%%%%%%%%%%%%%%%%%%%%%%%%%%%%%%%%%%%%%%%%%%%%%%%
\begin{figure}[t]
\begin{center}
\mbox{~\epsfig{file=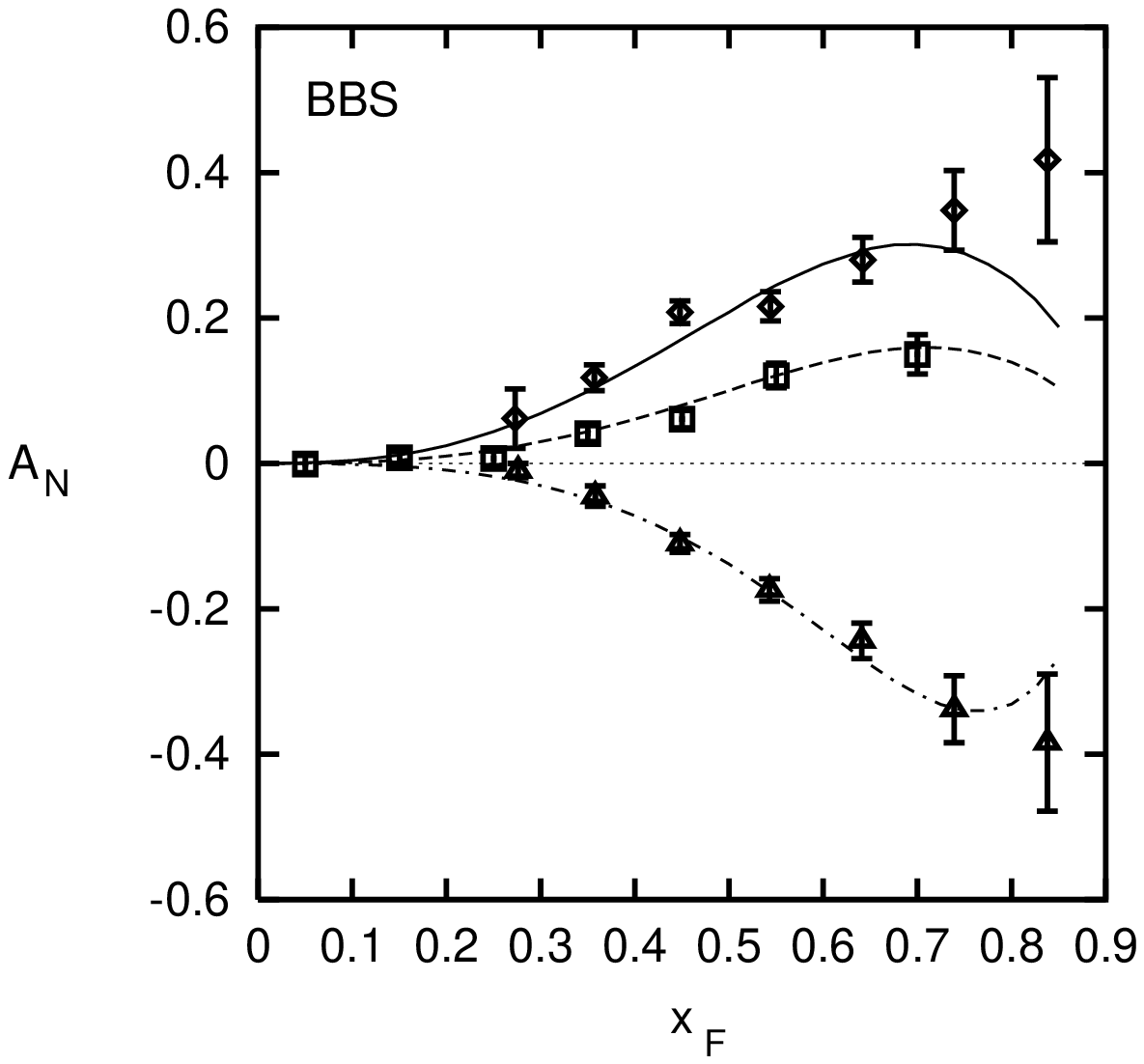,angle=-90,width=8.0cm}}
\vspace{0.4cm}
\caption{
The single spin asymmetry for pion production in the process 
$p^{\uparrow} p \to \pi X$ as a function of $x_F$, obtained by using the 
BBS \protect \cite{bbs} distribution functions. 
The solid line refers to $\pi^+$, the dashed line to 
$\pi^0$ and the dash-dotted line to $\pi^-$.}
\end{center}
\end{figure}
%%%%%%%%%%%%%%%%%%%%%%%%%%%%%%%%%%%%%%%%%%%%%%%%%%%%%%%%%%%%%%%%%%%%%%%%%%
%
%%%%%%%%%%%%%%%%%%%%%%%%%%%%%%%%%%%%%%%%%%%%%%%%%%%%%%%%%%%%%%%%%%%%%%%%%%
\begin{figure}[t]
\begin{center}
\mbox{~\epsfig{file=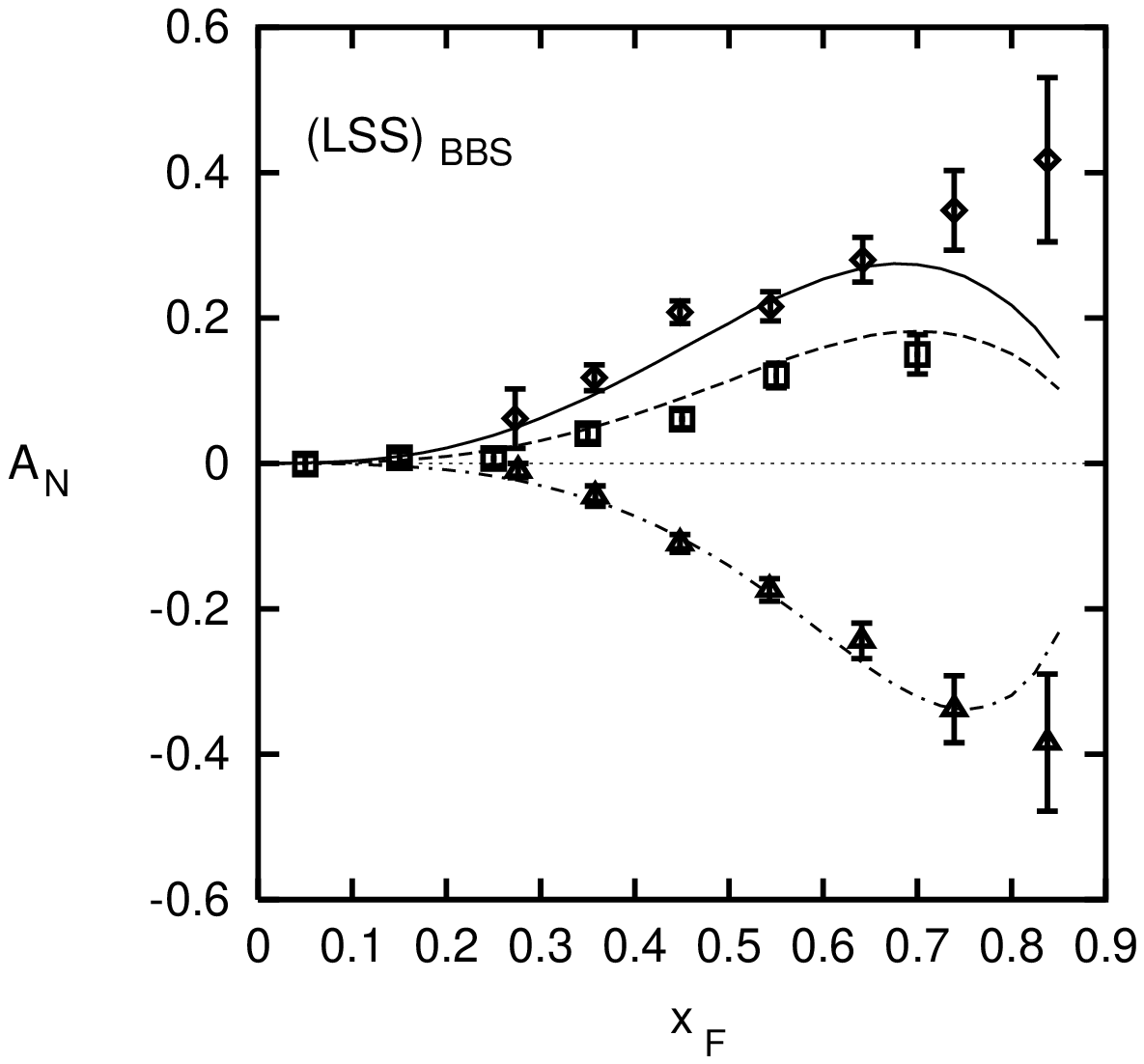,angle=-90,width=8.0cm}}
\vspace{0.4cm}
\caption{
The single spin asymmetry for pion production in the process 
$p^{\uparrow} p \to \pi X$ as a function of $x_F$,  determined by the fit 
using the (LSS)$_{BBS}$ \protect \cite{lss-bbs} distribution functions.
The solid line refers to $\pi^+$, the dashed line to 
$\pi^0$ and the dash-dotted line to $\pi^-$.}
\end{center}
\end{figure}
%%%%%%%%%%%%%%%%%%%%%%%%%%%%%%%%%%%%%%%%%%%%%%%%%%%%%%%%%%%%%%%%%%%%%%%%%%
%
%%%%%%%%%%%%%%%%%%%%%%%%%%%%%%%%%%%%%%%%%%%%%%%%%%%%%%%%%%%%%%%%%%%%%%%%%%
\begin{figure}[b]
\begin{center}
\mbox{~\epsfig{file=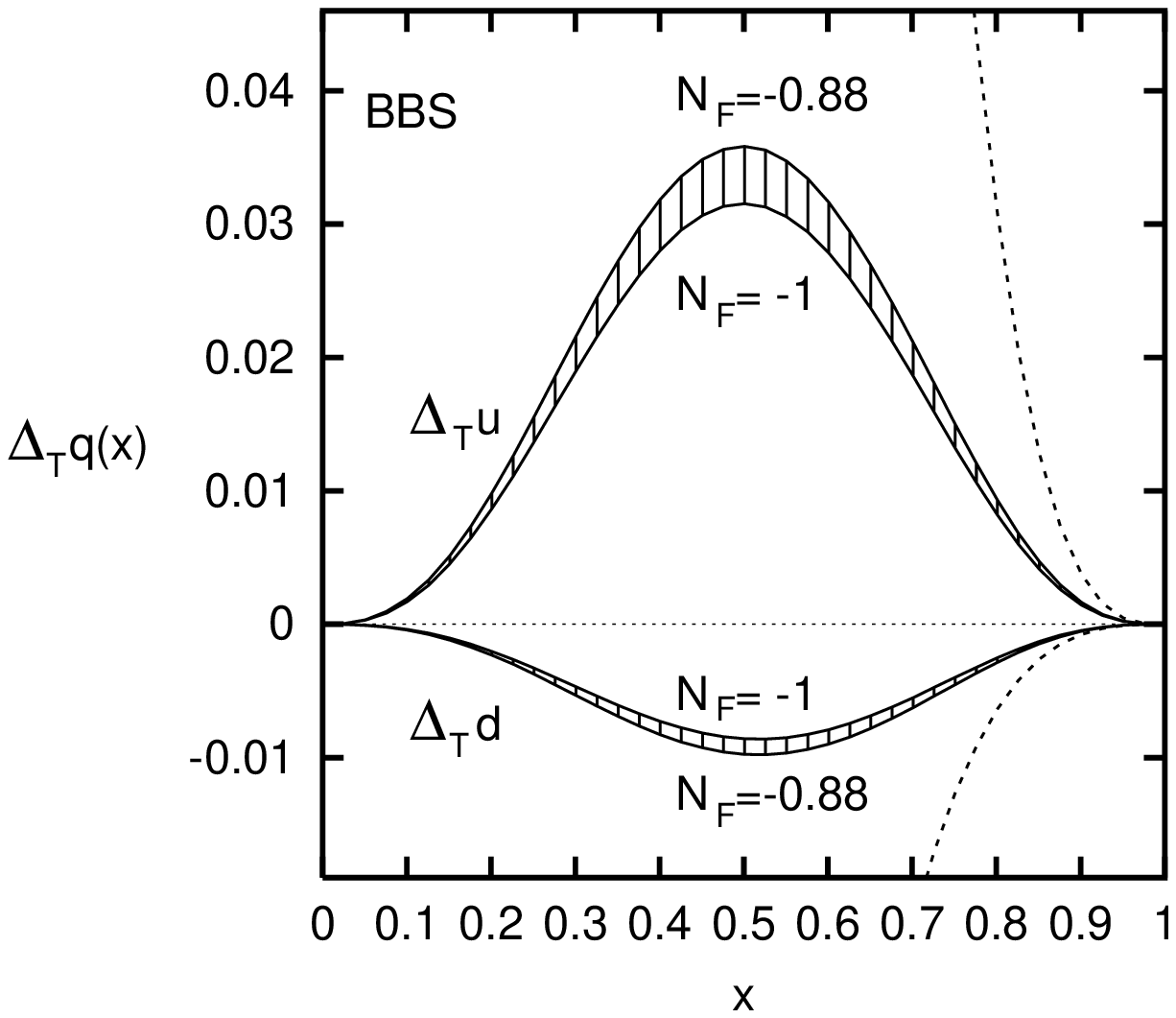,angle=-90,width=8.0cm}}
\vspace{0.4cm}
\caption{
The allowed range of distribution functions $\Delta _T u(x)$ and 
$\Delta _T d(x)$ versus $x$, 
as determined by the fit using the BBS \protect \cite{bbs} 
distribution functions.  The dotted lines are the 
boundaries imposed by the Soffer inequality.
For signs see discussion in text.}
\end{center}
\end{figure}
%%%%%%%%%%%%%%%%%%%%%%%%%%%%%%%%%%%%%%%%%%%%%%%%%%%%%%%%%%%%%%%%%%%%%%%%%%
%
%%%%%%%%%%%%%%%%%%%%%%%%%%%%%%%%%%%%%%%%%%%%%%%%%%%%%%%%%%%%%%%%%%%%%%%%%%
\begin{figure}[b]
\begin{center}
\mbox{~\epsfig{file=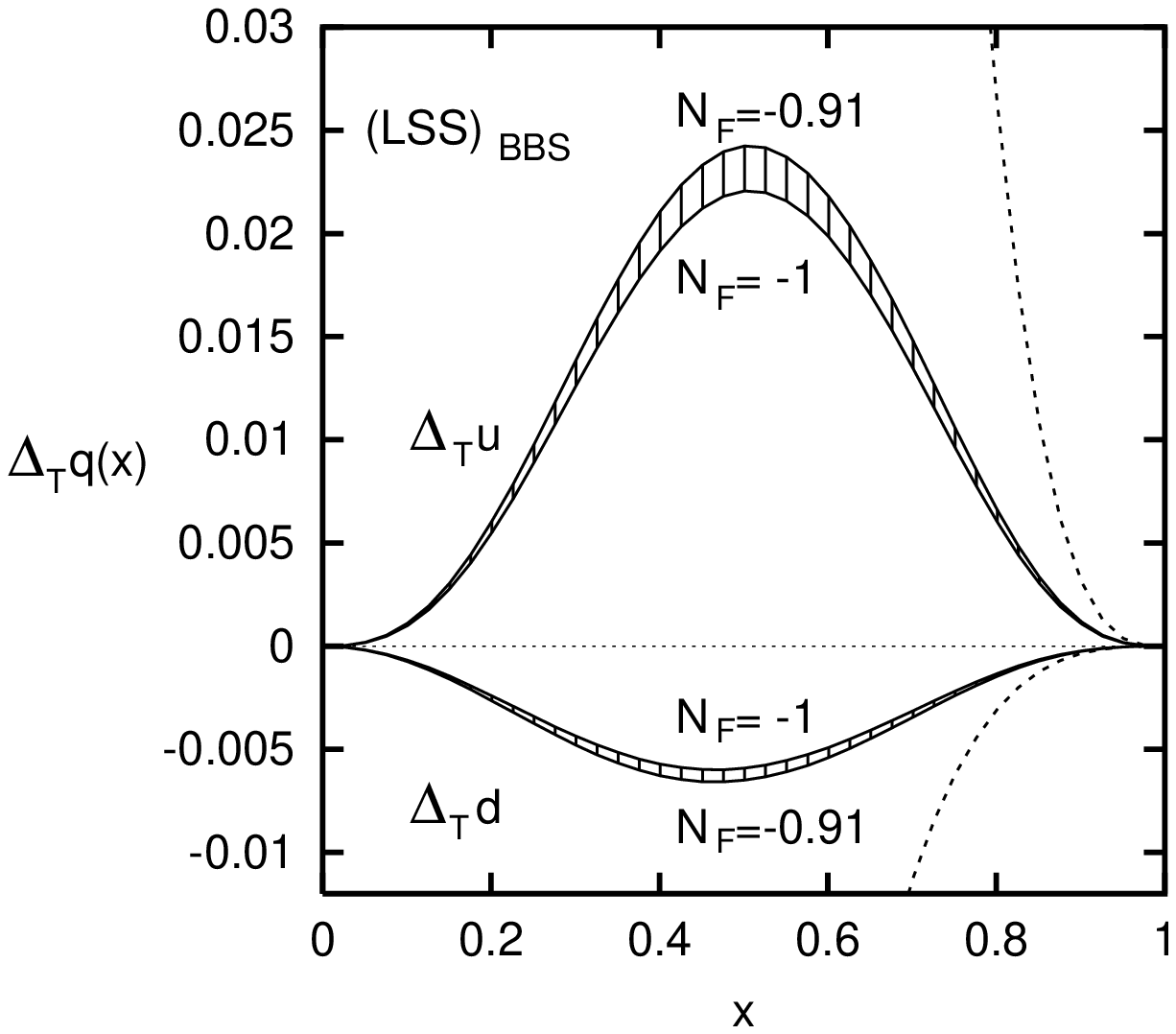,angle=-90,width=8.0cm}}
\vspace{0.4cm}
\caption{
The range of allowed distribution functions $\Delta _T u(x)$ and 
$\Delta _T d(x)$ versus $x$, 
as determined by the fit using the (LSS)$_{BBS}$ \protect \cite{lss-bbs} 
distribution functions. The dotted lines are the 
boundaries imposed by the Soffer inequality. For signs see discussion in text.}
\end{center}
\end{figure}
%%%%%%%%%%%%%%%%%%%%%%%%%%%%%%%%%%%%%%%%%%%%%%%%%%%%%%%%%%%%%%%%%%%%%%%%%%%
%
%%%%%%%%%%%%%%%%%%%%%%%%%%%%%%%%%%%%%%%%%%%%%%%%%%%%%%%%%%%%%%%%%%%%%%%%%%
\begin{figure}[ht]
\begin{center}
\mbox{~\epsfig{file=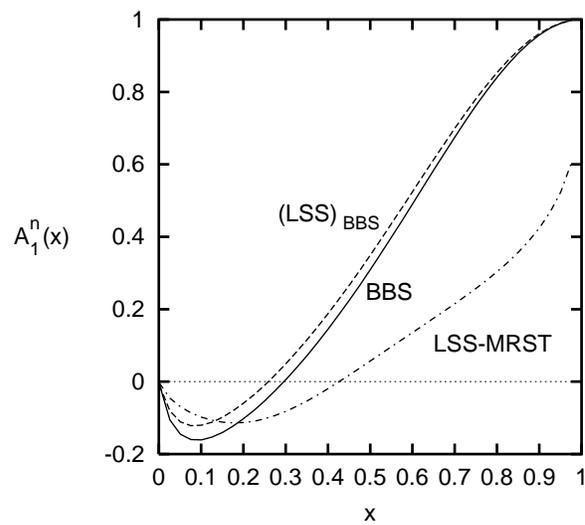,angle=-90,width=8.0cm}}
\vspace{0.4cm}
\caption{ The neutron longitudinal asymmetry $A_1^n(x)$ for $Q^2 \sim 1-4$ 
GeV$^2$, as obtained by using 
the BBS and (LSS)$_{BBS}$ parametrizations (solid and dashed lines 
respectively), and the LSS-MRST parametrizations (dash-dotted line).}
\end{center}
\end{figure}
%%%%%%%%%%%%%%%%%%%%%%%%%%%%%%%%%%%%%%%%%%%%%%%%%%%%%%%%%%%%%%%%%%%%%%%%%%
%
\end{document}